\documentclass[envcountsame,envcountsect]{llncs}

\usepackage[latin1]{inputenc} 
\usepackage{color}
\usepackage[normalem]{ulem}
\usepackage{graphicx}
\usepackage{wrapfig}

\begin{document}

\title{Geometric Theory for Program Testing}

\author{
Bernhard M\"oller\inst{1} \and
Tony Hoare\inst{2}  \and
Zhe Hou\inst{3} 
\and
Jin Song Dong\inst{3,4}
}

\institute{
Universit\"at Augsburg \and 
University of Cambridge and
Honorary Member of Griffith University \and 
Griffith University
\and 
National University of Singapore
}
\maketitle 

\begin{abstract}
Formal methods for verification of programs are extended to testing of 
programs. Their combination is intended to lead to benefits in reliable 
program 
development, testing, and evolution.
Our geometric theory of testing is intended to serve as the specification of a 
testing environment, included as the last stage of a toolchain that assists 
professional programmers, amateurs, and students of Computer Science.  The 
testing environment includes an automated algorithm which locates errors in a 
test that has been run, and assists in correcting them.  It does this by 
displaying, on a monitor screen, a stick diagram of causal chains in the 
execution of the program under test. The diagram can then be navigated 
backwards in the familiar style of a satnav following roads 
on a map.  This will reveal selections of places at which the program 
should be modified to remove the error.
\end{abstract}

\section{Introduction}

The aim of this work is to describe the design of a graphical display tool for inclusion as a component of an Integrated Software Development Environment (ISDE). Early versions of a partial ISDE are already delivering benefits in the Software industry. They reduce the lifetime costs of the development of significant commercial software products, particularly in the longest and last phase of their evolution, which typically lasts more than twenty years after initial launch. Our description of the design will exploit and develop your geometric intuition about the behaviour of a program when executed. 

Our ambition is that this will similarly develop the intuition of university students seeking a Degree of Master of Software Engineering. Without a professional education, few programmers will be found to use more advanced industrial ISDEs of the future.

The relevant formal methods for testing are due to the pioneers who provided 
the ideas: Euclid and Descartes for geometry; Carl Adam Petri, 
whose nets model execution of programs;  
Noam Chomsky, whose structured method defines the syntax of many  programming 
languages.  Their pioneering theories are simplified and adapted to meet 
current needs of programmers. 

A Euclidean diagram is formed by executing a 
set of constructors, whose feasibility is postulated by 
axioms and definitions. The geometric features of the diagram (axes, 
coordinates, points, lines, figures, ...) are labelled by identifiers chosen in 
drawing the diagram.  These identifiers relate the diagram to the proof of a 
Euclidean proposition, or the text of a program under test. 

As an example, we take a structured programming language, with program 
executions represented by Chomsky's Abstract Syntax Trees. A multiple 
simultaneous assignment labels the leaves of 
the tree with atomic commands, and constructors 
label the branching points.  Operators are sequential 
composition, object class declaration, and concurrent composition of 
various kinds.  Individual operations of the language are defined by 
specifying  the properties of a correct interface between their 
operands. Errors in arithmetic expressions can be detected by labelling a tree 
by the value that it produces. Detection of zero divide is then just a matter 
of calculation.  Other errors (eg. deadlock) can be defined by defining a 
pattern (eg. a cyclic chain of arrows).  

This makes it easy to define a new language feature separately by a new 
constructor.  A new language can be defined as the union of 
its features.  A testing tool should be automatically extensible to deal with 
any combination of features. 

\section{Preliminaries}

Our treatment of geometry owes everything to the ideas and inspiration of the pioneers of the subject. The Egyptian mathematician Euclid defined the basic concepts of plane geometry. He also invented a diagrammatic programming language, and its logic for proving correctness of its diagrams. We will exploit his treatment of points, lines, figures and their edges. We also exploit his famous fifth parallel postulate as the definition of parallel lines.

Two thousand years later, Descartes introduced orthogonal axes and coordinates, now shown on tablets of graph paper. He assumed that there is a single point as the unique element at the intersection of each pair of mutually orthogonal coordinates.  We will weaken this assumption by not requiring that every such intersection of coordinates should contain a point at all. In fact for our purposes of programming, most of the intersections will not.

In modern times, the US linguist Noam Chomsky introduced syntax trees to define the grammar of sentences of a natural language. These are now used widely in defining context free features of artificial languages as well.  A tree is defined recursively to be either a root or a branch or a leaf of a tree.  We will need only binary trees, a useful simplification in our diagrams.

\begin{figure}[ht!]
  \begin{center}
    \includegraphics[width=0.7\textwidth]{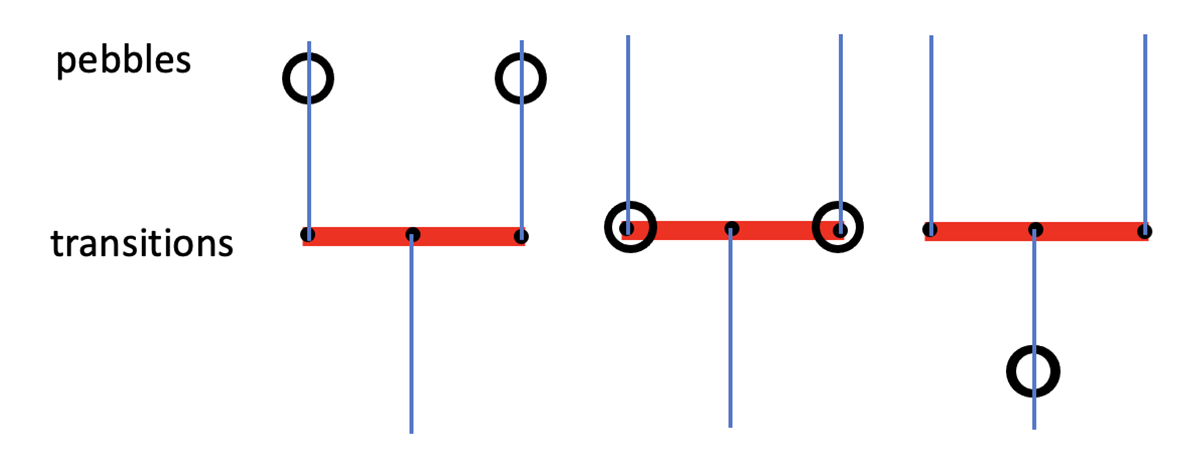}
  \end{center}
  \caption{An example of a Petri net transition.}
  \label{fig:petri-net}
\end{figure}

The German mathematician and computer scientist Carl Adam Petri introduced History Nets as dynamic logs (traces) of the behaviour of a computer in a single execution of a single program~\cite{petri2008}. A History Net consists of interconnected transitions. Each of these has a certain number of input and output arrows which may carry a single pebble or none. A transition can fire if and only if all its input arrows carry pebbles. The pebbles flow through the transitions from top to bottom like in Figure~\ref{fig:petri-net}.

\section{Geometric Diagrams}

Like the illustrations of the previous section, our directed graphs are diagrams, which can be drawn on a conventional sheet of graph paper. The nodes record events occurring in (or in a peripheral device of) a computer that is executing (or has executed) a program under test. The arrows represent a causal dependency relationship that prevents the event at its head from occurring before the event at its tail.  Hence the occurrence of the head event causally depends on the prior or simultaneous occurrence of the tail event. Hence the causation is an enabling condition that is necessary, not sufficient. For each event that occurs, it is necessary that all its enabling conditions have occurred, either earlier or simultaneously. 

As on conventional graph paper, there are two axes, horizontal space and vertical for time. The points represent all events occurring in the execution of a single node in the syntax diagram of the program text. 
In one of the earlier versions of this work, all lines are drawn as arrows. To reduce clutter, we now remove arrowheads and only draw lines, but we may still use the words \emph{arrow} and \emph{line} interchangeably throughout this work. 

\begin{figure}[ht!]
  \begin{center}
    \includegraphics[width=0.8\textwidth]{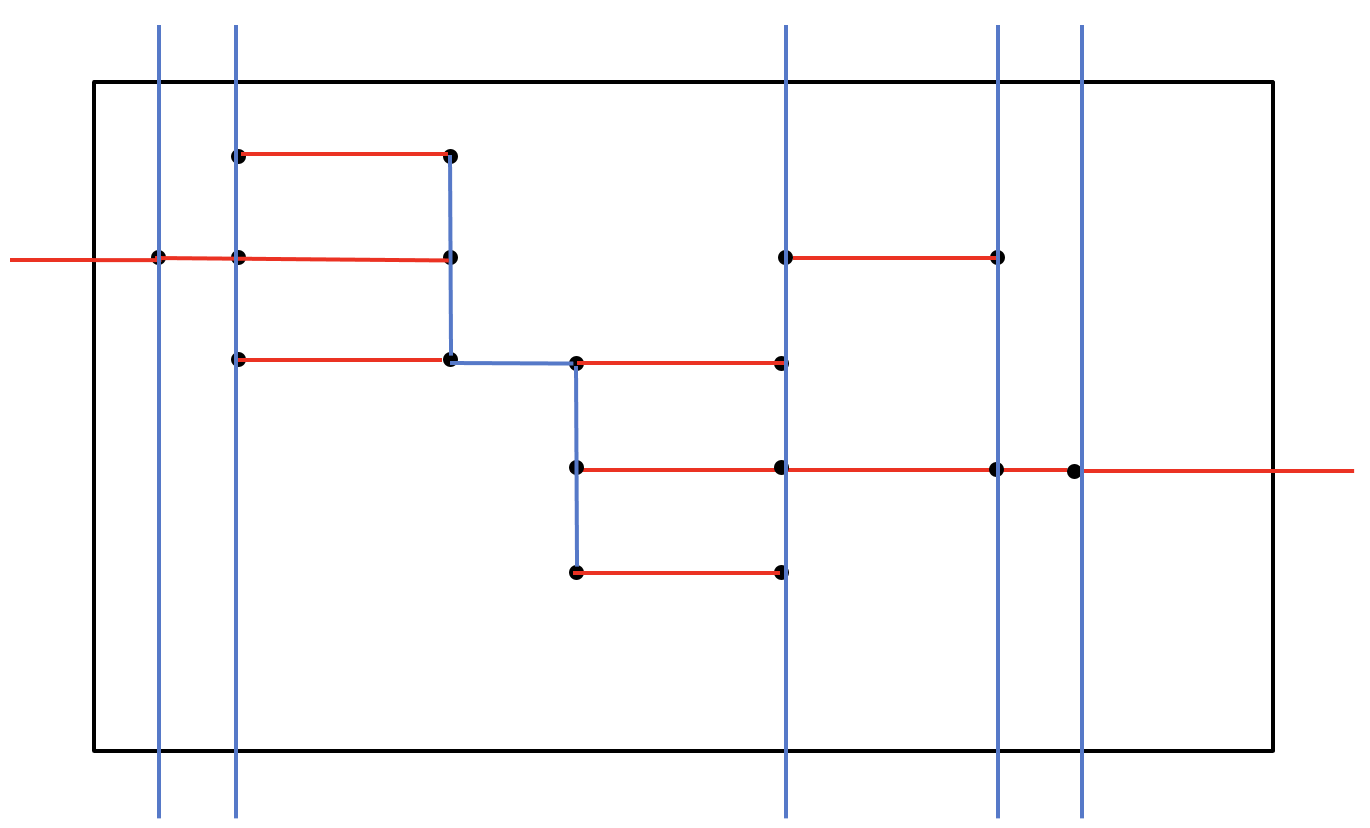}
  \end{center}
  \caption{An example of a geometric diagram that captures objects, communication, and atomic actions.}
  \label{fig:diagram}
\end{figure}

Figure~\ref{fig:diagram} shows seven blue vertical lines, drawn along the vertical coordinates of the graph. They are mutually parallel in the Euclidean sense that they do not share any points. Each of them records the complete history of the actions of a single separate object in computer memory, for example, a resource, a variable, a channel or a thread of a concurrent program. The points that lie on each coordinate represent the successive observations performed by or on that object. So implicitly all blue arrows point downwards. 

The local coordinates of a slice are those that do not pass through any of the surrounding edges of the slice, as shown in the middle of this diagram. The earliest point of such a coordinate is the allocation of a resource, and the bottom coordinate is its disposal. This remark gives an intuitive definition of the purpose and meaning of allocation and deletion. It gives no hint of how the purpose may be achieved, at compile time or in a stack or on the heap.  Nor does it determine where or when each event occurs.  This is decided by the implementation of the language, usually at run time.
 
The other five vertical lines extend through the edges of the slice, beyond the slice, both above it and below. They are global resources owned throughout this execution. This shows that the horizontal coordinates continue into the neighbourhood on the left and above to reach the horizontally neighbouring slice. The vertical coordinates do the same for the neighbourhoods above and below. 

The two new red lines at the edges of this diagram are synchronised communications inputting to the diagram on the left, and outputting to the right.

The blue horizontal line in the middle of the diagram represents a transfer of ownership from one thread to another. It shows that release and acquisition of a resource are synchronised events. 

A horizontal coordinate is drawn with red arrows. It records the virtually simultaneous execution of points lying on it. These multiple points may be terminal symbols of the grammar, corresponding to basic commands, or they may be method calls to class declarations at a lower level of abstraction.  The points on each horizontal coordinate represent the simultaneous performance of events by two or more objects. It must include a thread that issued the command, and it must be an action of an object that is currently owned by that thread. The coordinate satisfies the classical definition of an atomic action, namely that there does not exist any instant in time at which some of its actions have been observed to occur, and others have not. Again, the method of implementation and program execution is not mentioned. 
     
There may be various roles that horizontal coordinates play in an application.  For example, they can be commercial transactions in blockchains, or fences in relaxed memory.

\section{Labels in the Diagrams} 

We have described the order of execution of events of the slice, but we have not given any meaning to the internal components of the program.  For this we attach symbolic names to the components, coordinates, and basic commands. Each name is one that appeared originally in the source program.  This is essential in program testing, to understand where and how the program should be corrected. 

The purpose of a label is to correlate a diagram with the text of the segment of a program whose execution it displays.
The label on a vertical coordinate is an object name. 
On a horizontal coordinate, it is a basic command or method call. 
Each arrow may also be labelled by a relevant value computed and used during program execution.

\begin{figure}[ht!]
  \begin{center}
    \includegraphics[width=0.9\textwidth]{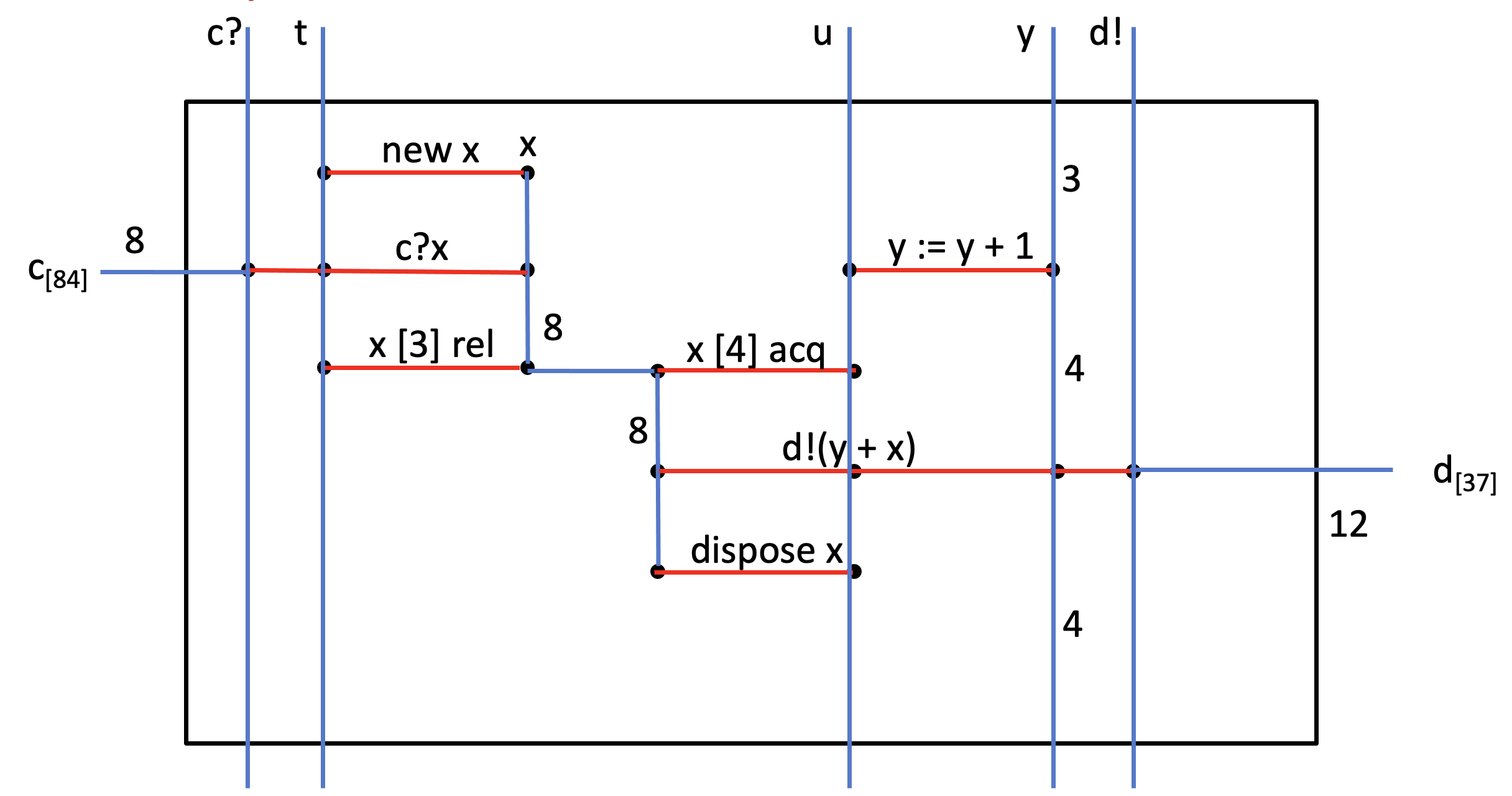}
  \end{center}
  \caption{An example of a labelled diagram that captures variables, threads, input, output, allocation, disposal, and ownership transfer.}
  \label{fig:labelled-diagram}
\end{figure}

We give an example labelled diagram in Figure~\ref{fig:labelled-diagram}.

The input and output interface coordinates are labelled by the unique name or address of the object whose behaviour it records. Each name belongs implicitly to some object class, for example, an output thread named $d!$, or a variable named $y$, or an output port for $d$ on the subscripted name  $d_{[37]}$ of the 37th message communicated on the channel $d$. The class defines and determines the correct behaviour of all its objects. 

The left part of Figure~\ref{fig:labelled-diagram} has two thread names, $t$ and $u$.
The name $x$ uniquely identifies an integer variable allocated and disposed locally to this diagram. It is represented by the location number (i.e., reference) of the object in computer memory. The label $c?$ denotes the input port of a communication channel $c$. 

Each horizontal coordinate of the diagram is labelled by the basic command in the thread, which has triggered the execution of the atomic event. For example, on the left of this slide, the actions of the thread $t$ are an allocation of a new object $x$; an input of a message on channel $c$ simultaneously with its assignment to $x$; and finally a release of ownership of $x$. 

On the right of Figure~\ref{fig:labelled-diagram} there are the actions of the thread $u$. They are an incrementation of the variable $y$, an acquisition of the ownership of $x$, an output on channel $d$, and a disposal of the object $x$. The indices on the release and acquire commands give the serial numbers of these actions in the life of the variable $x$. 

The numeric labels on each arrow indicate the value transmitted from the tail to the head of the arrow by the performance of both these actions. On a vertical arrow, the label is the current value of the object during the interval between the occurrence of the tail action and of the head action. On a horizontal coordinate arrow, it is the value of the message communicated. The correctness of the values may be easily checked against the rules that define each expression or basic command in the language. 

The state of the executing computer memory at the beginning of execution of a slice (and at the end) is defined as the function that maps the name of each arrow crossing the top (or bottom) edge to the value which labels the arrow. 

For example, at the top edge of Figure~\ref{fig:labelled-diagram}, the initial state maps the variable named $y$ to the value $3$, whereas at the bottom edge the final state maps the variable $y$ to the value $4$. A similar function describes all the interactions between neighbouring slices that occurred during execution. For example, $c_{[84]}$ has the value $8$. 

Our model of program execution so far has been entirely event-based. We have just explained how memory can be introduced into an event-based model: it is defined in terms of events, by exploiting the labels for objects and values that are attached to the vertical coordinates. 

As an overview, let us look at the whole story of part of the execution of an extremely unlikely segment of a program. The story starts at the top left with the allocation of a new object $x$ by the thread $t$. This is followed by input of the initial value $8$ from its local input port $c?$. This is simultaneously assigned to $x$. The thread $t$ then releases ownership of $x$ and $u$ simultaneously acquires it. Meanwhile, on the right side of Figure~\ref{fig:labelled-diagram}, the thread $u$ has incremented the value of its own local variable $y$. It then acquires $x$ from the thread $t$, and outputs on its output port $d!$ the sum of $x$ and $y$. Finally it disposes the object $x$. 

\begin{figure}[ht!]
  \begin{center}
    \includegraphics[width=0.5\textwidth]{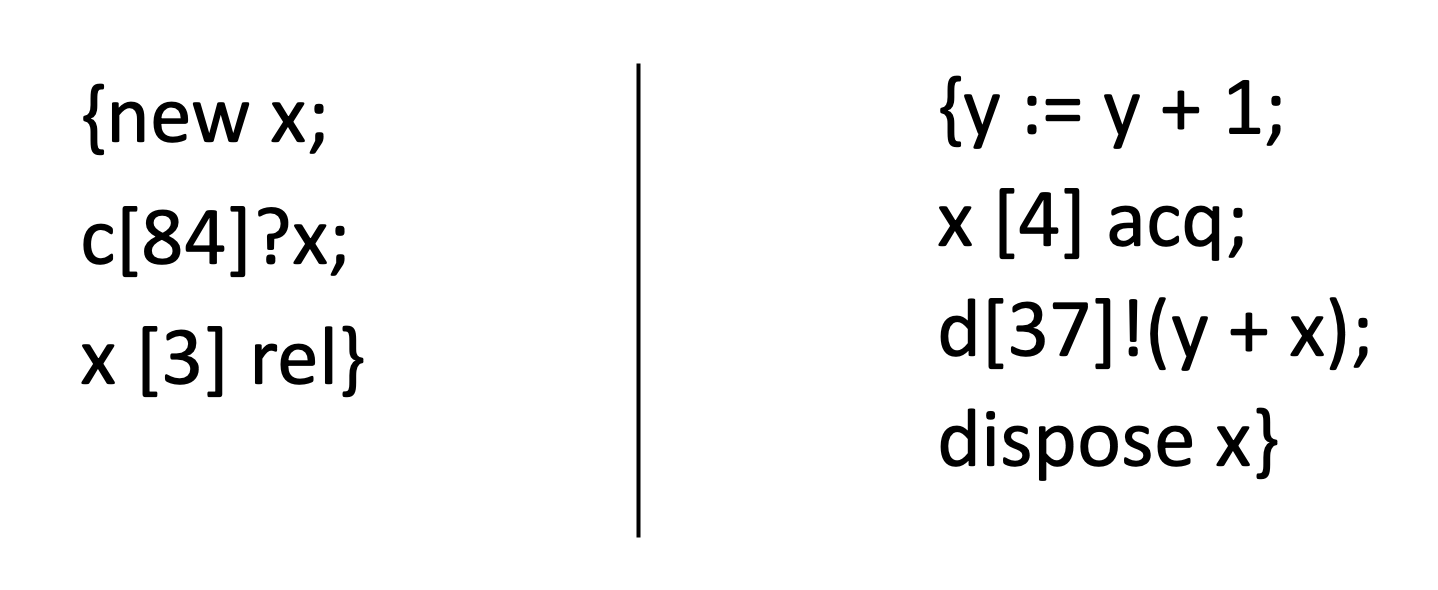}
  \end{center}
  \caption{An example of program code that corresponds to Figure~\ref{fig:labelled-diagram}.}
  \label{fig:code}
\end{figure}

Figure~\ref{fig:code} tells the same story as Figure~\ref{fig:labelled-diagram}, expressed in the form of the text of the segment of program which was executed. It uses semicolon in a familiar way to denote sequential composition, and the single vertical bar denotes concurrent composition, as in process algebras like Communication Sequential Processes (CSP)~\cite{hoare1978}.

It is a great simplification to treat concurrent composition as a simple binary program connective, just like the semicolon which calls for sequential execution. Their properties, like transitivity are expressed axiomatically, and the program can be transformed for purposes of compilation or optimisation.

\section{Definition of Operators}

We are now ready to add program operators to the diagram. We will look at sequential and concurrent compositions, and their combinations.

The definitions of the composition operators of a programming language follow the normal style of definitions in the natural sciences and in applied mathematics. Each technical term is defined by relating it to the real-world phenomena which it denotes.  This often describes in general terms the purpose and intended effect of the defined concept, without giving even a hint of how it is implemented. 

\begin{figure}[ht!]
  \begin{center}
    \includegraphics[width=0.8\textwidth]{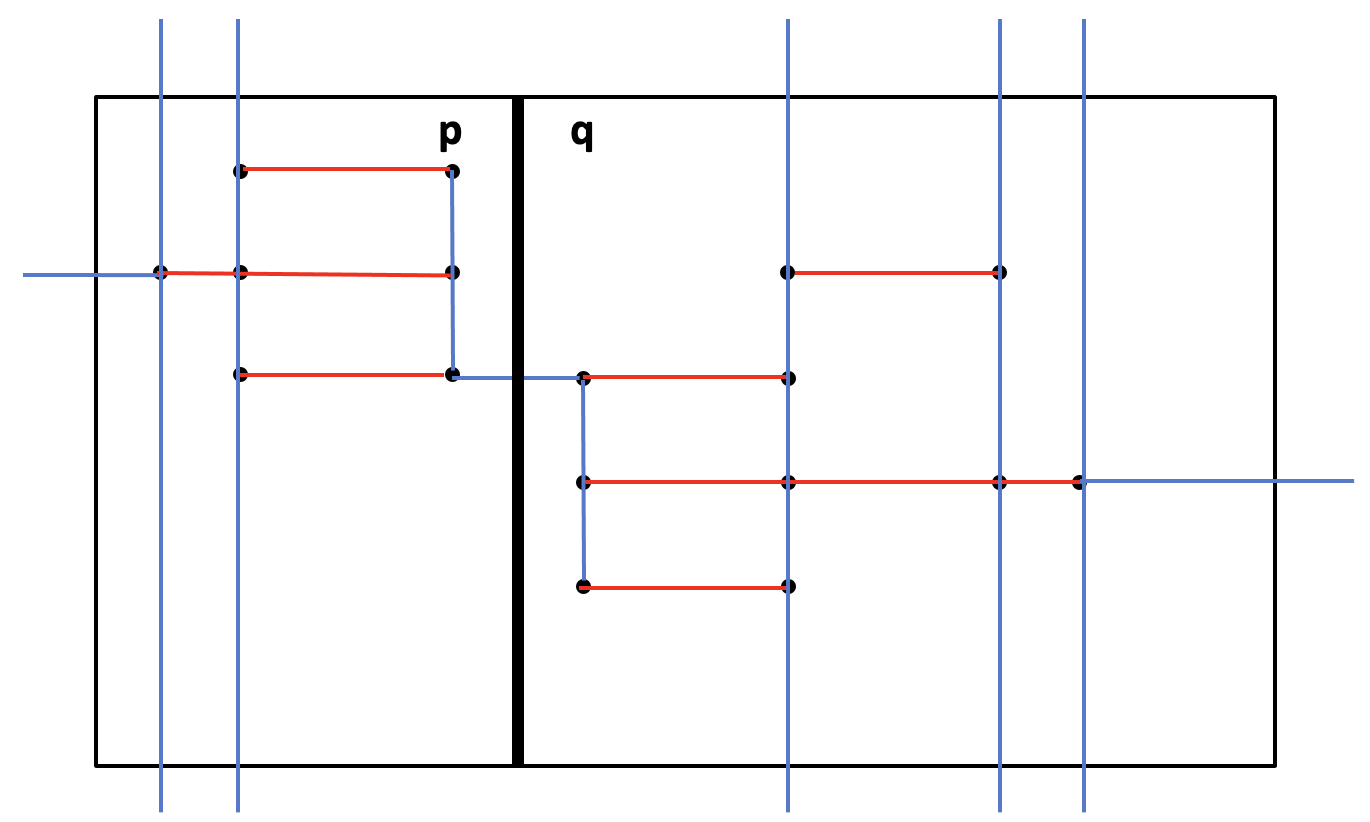}
  \end{center}
  \caption{An example of concurrent composition in diagram.}
  \label{fig:concurrent}
\end{figure}

Figure~\ref{fig:concurrent} shows a vertical slicing of a diagram by introducing an internal vertical edge. $p | q$ gives a concrete syntax for the segment of program that was executed. The vertical bar symbol indicates concurrent composition of its two operands. The new splitting edge (together with all the arrows which cross it) is shared between the two component boxes. It appears as the right vertical edge of $p$ and as the left vertical edge of $q$. They are indeed the same edge! The horizontal coordinates are also split into disjoint line slices at the points where the newly introduced edge crosses them. 

It is immediately obvious from the horizontal edge at the top of the diagram that the two components p and $q$ start together and finish together, as required by the definition of the purpose of concurrent composition. Furthermore, the top edge of $p$ is wholly disjoint from the top edge of $q$. Consequently, the vertical coordinates that cross the top edge of $p$ are wholly distinct from those that cross the top edge of $q$. The same is true of the bottom edges. This disjointness is the intuition behind Peter O'Hearn's separation logic~\cite{ohearn2019}. The local variables of a slice are called its footprint.

An ownership discipline for threads and objects is designed to prevent a race between two threads for access to the same object. Such a race is a source of uncontrollable non-determinism.  We have already prohibited races by the rule that every vertical edge between two threads must pass through some horizontal blue arrow. 
Synchronisation of the transmission of ownership is generally more efficient than general synchronisation for horizontal coordinates.  Like sequential composition, it often has no run-time overhead.  

\begin{figure}[ht!]
  \begin{center}
    \includegraphics[width=0.8\textwidth]{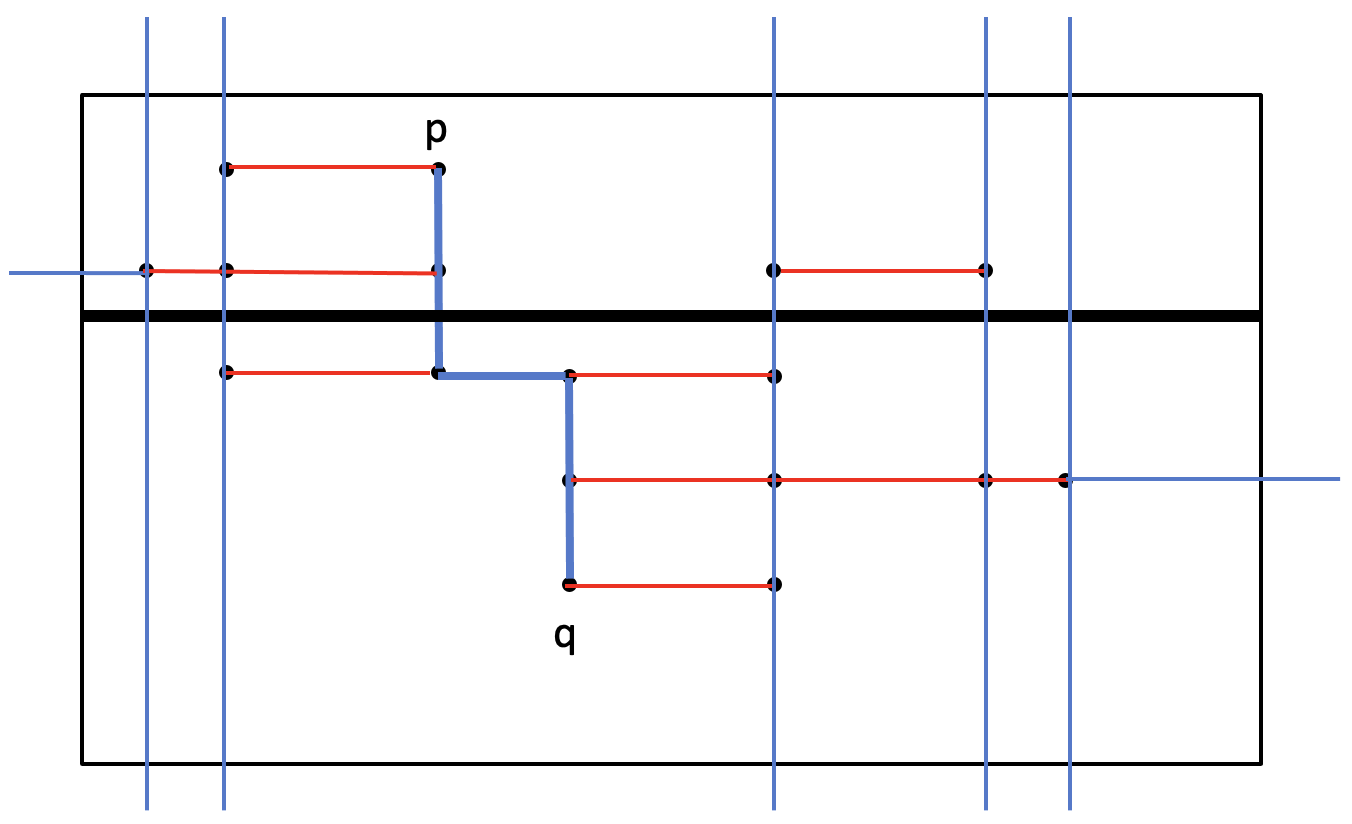}
  \end{center}
  \caption{An example of sequential composition in diagram.}
  \label{fig:sequential}
\end{figure}

The semicolon denoting sequential composition is represented geometrically by a horizontal split in Figure~\ref{fig:sequential}. The state of memory of the executing machine passed from $p$ to $q$ is defined by the edge which they share, in the manner that we have described earlier.

Note that it is geometrically impossible for any arrow to point upwards across a horizontal split from $q$ to $p$. If such a crossing were possible, the principle of causation would require the implementation to start executing the component $q$ before finishing the execution of $p$. This is ruled out by the definition of semicolon together with the definition of causation. 

In Euclidean plane geometry, it is impossible also for a horizontal arrow to cross a horizontal coordinate, which is intended to denote an atomic event. Thus the memory can never be in a state where some of the actions on the coordinate have already happened, and some of them have not. It is prohibited as our definition of parallelism by Euclid's famous but controversial parallel postulate that forbids it. 

\begin{figure}[ht!]
  \begin{center}
    \includegraphics[width=0.8\textwidth]{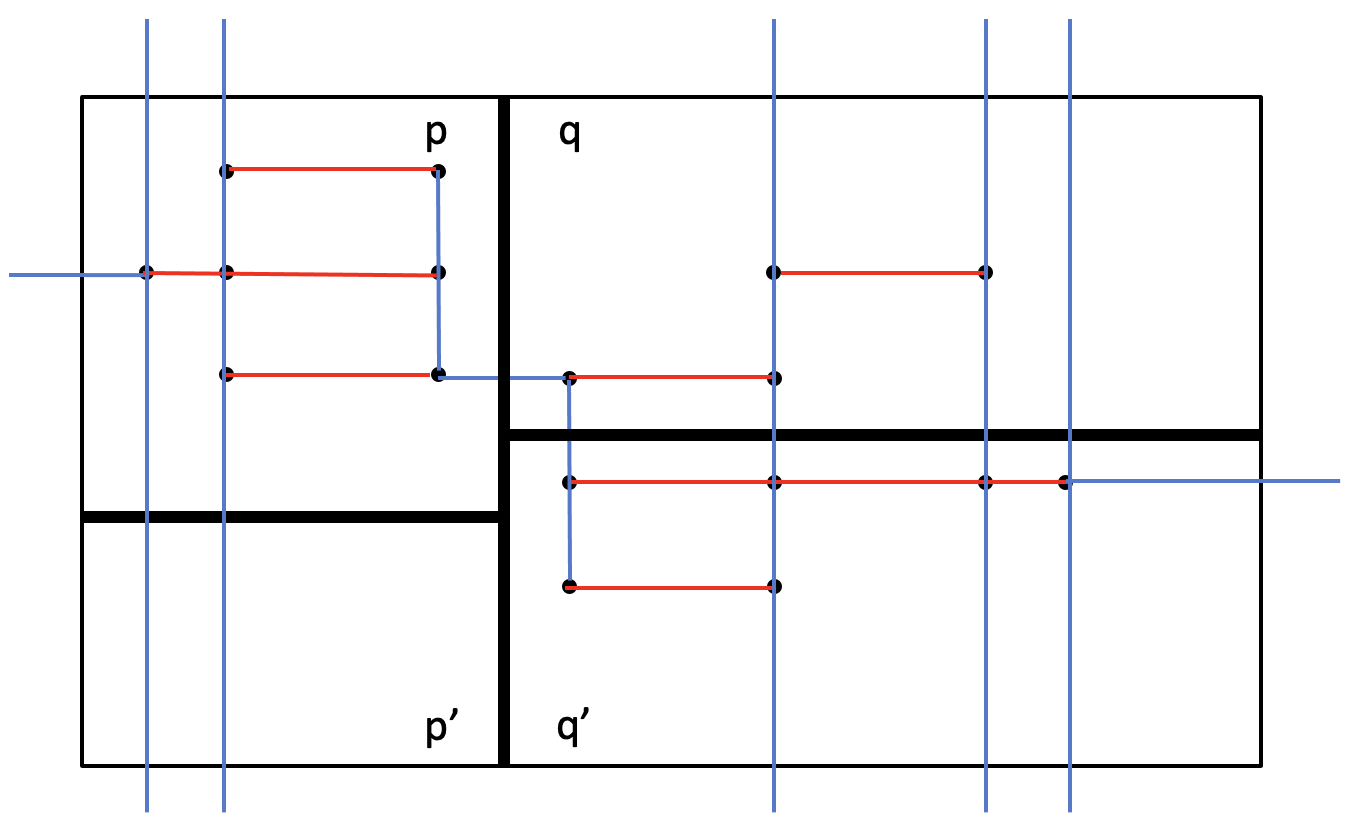}
  \end{center}
  \caption{A combination of sequential and concurrent compositions in diagram.}
  \label{fig:combo1}
\end{figure}

\begin{figure}[ht!]
  \begin{center}
    \includegraphics[width=0.8\textwidth]{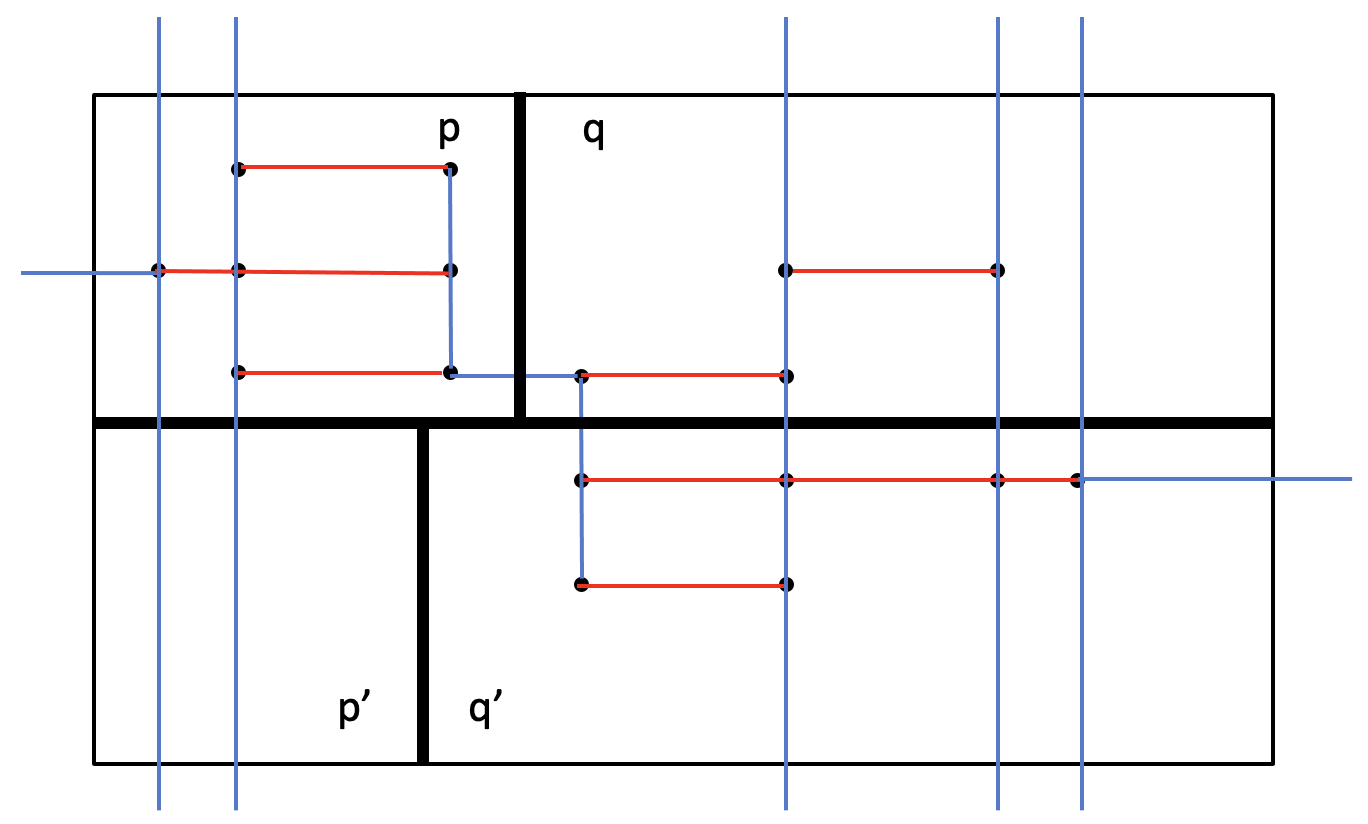}
  \end{center}
  \caption{Another combination of sequential and concurrent compositions in diagram.}
  \label{fig:combo2}
\end{figure}

Figure~\ref{fig:combo1} illustrates how diagrams formed by a previous split can be split again. The bracketed operator of the program gives the orientation of the first split. Subsequent splits on the internal slices may be either parallel or orthogonal to it. The process of splitting may be continued until all the diagrams contain only a single horizontal coordinate.
That is how Figure~\ref{fig:combo1} is distinguishable from Figure~\ref{fig:combo2}, in which the first split is horizontal, and the second and third splits are vertical. 

In summary, we have illustrated the geometric constructions by application to an example program execution, which includes many of the essential features of C++ and other modern object-oriented languages. They are listed below. They include features that have been omitted from most formalisations of programming language designs. In turn, we omit features that are common to nearly all such formalisations. 

\begin{itemize}
    \item Objects: variables, channels, threads, other resources.
    \item Transactions: aka fences, atomic events, Petri net transitions.
    \item Allocation, disposal and transfer of ownership of resources.
    \item Variables (in stack or heap), synchronous channels, threads.
    \item Assignments updating to variables by threads.
    \item Communications through input and output ports.
    \item Sequential and concurrent composition, choice.
    \item \textbf{Omitted}: types, conditionals, loops, timing, waiting, probability.
\end{itemize}

As in the Elements of Euclid, geometric diagrams are good for conveying ideas and intuitions. The only formal text material in the diagrams will be labels that name the components of the diagram.  They are the same sort of names, both in Euclid's proofs and in our programs.

\section{Testing Environment}

Let us conclude with a discussion on program testing. The purpose of testing is 
\begin{itemize}
    \item to locate errors in the program under test --- particularly errors in the interfaces between large components and those in recently changed code;
    \item to assist in their diagnosis and the search for a correction or an acceptable work-around;
    \item to record evidence for code reviews and analysis.
\end{itemize}

Any realistic theory for program testing must be based on a study of the various kinds of errors that might be detected in the execution of an arbitrary program. These may be classified in two ways: as errors in the program under test, which the programmer is capable of mending, or as errors that require mending by someone else. We include errors that can only be due to a fault in the implementation itself below. The first two points violate rules which we have quoted as part of the very definition of the programming language. 

\begin{itemize}
    \item An arrow that crosses from $q$ to $p$ in $(p ; q)$.
    \item A horizontal arrow that crosses a vertical coordinate.
    \item A failure to detect falsity of an assumption (precondition).
    \item An error that should have prevented execution.
    \begin{itemize}
        \item A syntactic error.
        \item A strict type mismatch.
        \item A command that is not syntactically in the programming language.
    \end{itemize}
\end{itemize}

There are also generic programming errors that are explicitly ascribed to the program itself. They are the errors that program testing is aimed at detecting and are usually attributed to the program developer. Examples include
\begin{itemize}
    \item non-termination (including deadlock),
    \item a vertical split crossing a red horizontal arrow, which violates the atomic action discipline;
    \item an operation that is explicitly forbidden by the language standard such as zero divide, null dereference, subscript overflow $\ldots$
    \item an assertion that evaluated to false, which violates a promise made by the programmer.
\end{itemize}

It is easy to introduce new operators into our geometric model of program execution, by simply placing different restrictions on the kinds of arrows that cross between its operands. In the definitions below, the first one with the triple vertical bards is the most restrictive, and the second one with two greater than signs is least restrictive. Thus they also rule out deadlock. 

\begin{itemize}
    \item $p~|||~q$ (interleaving in CSP or separating conjunction in separation logic) forbids any arrow crossing between $p$ and $q$.
    \item $p >\!\!> q$ (chaining or a pipe in Unix) forbids any arrow crossing from $q$ to $p$.
\end{itemize}

So much for the theories, but in an ideal world we would like to see the theories being implemented in user-friendly tools. Such a tool should be able to select and/or generate a suite of test cases that exercise recently changed code; locate all violations in the resulting traces, and mark the responsible commands in the program listing; present a navigable display of the trace allowing travel along causal chains in either direction; Zoom on multiple selected regions of the diagram, etc.

This work is extracted from the keynote prepared by Hoare with the assistance of M\"oller and Hou for the ATVA 2021 conference~\cite{atva2021}. It is part of a series of related lectures, monographs, and summer schools listed below.  
\begin{itemize}
    \item Algebraic theory for program design (and optimisation).
    \item Predicate calculus for program specification (and proof).
    \item Quantum theory for quantum programming (and machine learning).
    \item Navigation tools for error detection (and repair).
\end{itemize}

Much of related work is done by colleagues in the Verified Software Initiative, which was started as a wide international collaboration in 2005. Its second of three phases will build on its achievements in the construction of Theories Tools and experiments. Anyone of the readers will be welcomed to join it.

\bibliographystyle{plain}
\bibliography{main.bib}

\end{document}